\documentclass[sigconf]{acmart}

\makeatletter
\renewcommand\footnotetextcopyrightpermission[1]{}
\makeatother
\acmConference{}{}{}
\acmBooktitle{}

\acmISBN{}
\acmDOI{}
\usepackage{enumitem}

\AtBeginDocument{%
  }

\usepackage{algorithm}
\usepackage{algpseudocode}
\usepackage{pifont}
\usepackage{xcolor}

\usepackage{subcaption}
\usepackage{tikz}
\newcommand{\redcirc}[1]{%
\tikz[baseline=(char.base)]{
    \node[shape=circle,draw=red,inner sep=1pt,text=red] (char) {#1};
}}
\newcommand{\LComment}[1]{%
  \Statex \textcolor{blue}{$\triangleright$\ \textit{#1}}%
}

\newcommand{\LCommentbf}[1]{%
  \Statex \textcolor{blue}{$\triangleright$\ \textbf{#1}}%
}

\newcommand{\LCommenttab}[1]{%
  \Statex \hspace{1.5em}\textcolor{blue}{$\triangleright$\ \textit{#1}}%
}

\newcommand{\LCommenttabtab}[1]{%
  \Statex \hspace{3em}\textcolor{blue}{$\triangleright$\ \textit{#1}}%
}

\begin{document}

\title{RSH-SpMM: A Row-Structured Hybrid Kernel for Sparse Matrix-Matrix Multiplication on GPUs }
\renewcommand{\shorttitle}{ }
\author{Aiying Li}
\email{lay_sn1987a@mail.ustc.edu.cn}

\affiliation{%
  \institution{University of Science and Technology of China}
  \city{Hefei}
  \state{Anhui}
  \country{China}
}

\author{Jingwei Sun}
\email{sunjw@ustc.edu.cn}

\affiliation{%
  \institution{University of Science and Technology of China}
  \city{Hefei}
  \state{Anhui}
  \country{China}
}

\author{Han Li}
\email{hanli06@mail.ustc.edu.cn}

\affiliation{%
  \institution{University of Science and Technology of China}
  \city{Hefei}
  \state{Anhui}
  \country{China}
}

\author{Wence Ji}
\email{jiwence@mail.ustc.edu.cn}

\affiliation{%
  \institution{University of Science and Technology of China}
  \city{Hefei}
  \state{Anhui}
  \country{China}
}

\author{Guangzhong Sun}
\email{gzsun@ustc.edu.cn}

\affiliation{%
  \institution{University of Science and Technology of China}
  \city{Hefei}
  \state{Anhui}
  \country{China}
}

\begin{abstract}
 Sparse Matrix-Matrix Multiplication (SpMM) is a fundamental computation in graph analytics, scientific simulation, and sparse deep learning workloads. However, the extreme irregularity of real-world sparse matrices prevents existing GPU-based methods from maintaining high Tensor Core utilization and stable throughput. We present \textbf{RSH-SpMM}, a fine-grained row-structured hybrid SpMM framework designed to better align irregular sparsity with modern GPU execution pipelines. RSH-SpMM introduces adaptive row partitioning and employs the RS-Tile representation to expose Tensor-Core-efficient dense fragments, while processing irregular rows on a minimal-overhead CUDA execution path. It further employs a load-balanced hybrid kernel with locality-aware reordering to enhance structural coherence and sustain high execution efficiency under highly irregular sparsity. Comprehensive evaluations across diverse sparse workloads demonstrate that RSH-SpMM consistently outperforms state-of-the-art SpMM designs, yielding \textbf{1.27$\times$--6.13$\times$} acceleration and maintaining robust performance across matrices with highly irregular sparsity structures.
\end{abstract}

\maketitle

\section{Introduction}

Sparse Matrix-Matrix Multiplication (SpMM) is a core computational primitive in Graph Neural Networks (GNNs)~\cite{computinggnn,maxkgnn,pckgnn,gnnpilot}, graph analytics~\cite{pagerank,graphblast}, scientific computing~\cite{VAZQUEZ2010146,MatrixKrylov}, and modern sparsity-aware Large Language Model (LLM) inference~\cite{sparsegpt,wanda,flashllm}. Despite its ubiquity, attaining high GPU performance remains difficult because real-world sparse matrices exhibit extreme structural heterogeneity: heavy-tailed row-length distributions, rapidly shifting local densities, and fragmented nonzero patterns. Such irregularity disrupts warp-level parallelism, limits memory coalescing, and is fundamentally misaligned with the dense, tile-oriented execution model required for efficient Tensor Core (TC) utilization~\cite{bell2009spmv,graphblast}.

\begin{figure}[t]
  \centering
  \includegraphics[width=0.95\linewidth]{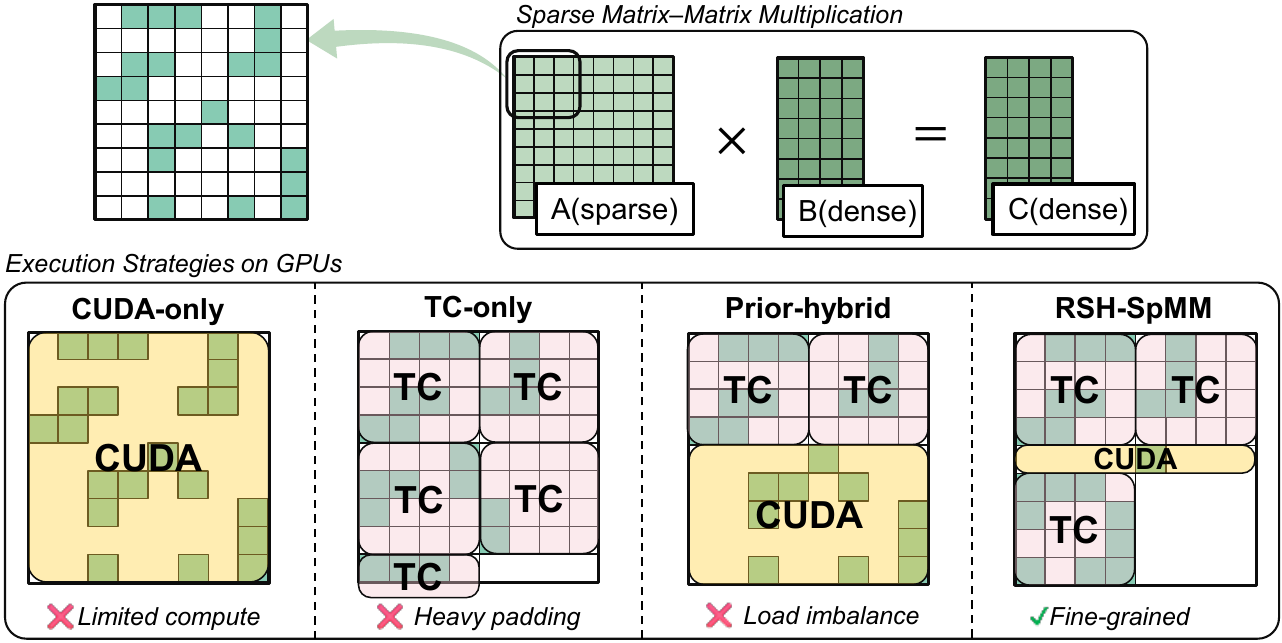}
  \caption{Comparison of GPU SpMM execution paradigms. }
   \label{fig:intro}
\end{figure}

Modern GPUs expose two complementary compute units: CUDA cores, which offer relatively flexible and fine-grained execution for irregular sparsity, and Tensor Cores, which provide much higher throughput but require dense, tile-aligned operands. Existing GPU SpMM methods fall into three categories, as illustrated in Figure~\ref{fig:intro}.
\emph{CUDA-core approaches} provide robust handling of irregular sparsity through warp-cooperative processing~\cite{Heuristic,sputnik,ge-spmm}, yet remain bounded by scalar arithmetic throughput and cannot exploit the substantial performance advantage offered by Tensor Cores.
\emph{Tensor-Core approaches} reorganize sparse matrices into dense-aligned blocks through windowing schemes, bitmap compression, or fragment reshaping~\cite{tc-gnn,dtc-spmm,acc-spmm}, yet their fixed window sizes, rigid fragment geometries, and uniform tiling rules yield low tile density and unstable performance when local sparsity varies.
\emph{Hybrid approaches} attempt to route locally dense regions to Tensor Cores while directing irregular regions to CUDA cores~\cite{hc-spmm,brp-spmm}. However, their partitioning strategies typically operate at coarse granularity, making decisions at the matrix level or across large block regions, and fail to capture fine-grained structural coherence within and across row groups, leaving substantial opportunities for Tensor-Core acceleration unexploited.

These limitations point to a key gap: existing methods lack a \emph{fine-grained, structure-aware execution strategy} that operates at the granularity where sparsity naturally varies. An effective solution should identify row groups suitable for Tensor Core execution and handle irregular rows with CUDA-core kernels. This approach enables compact storage, lightweight decoding, and stable performance under diverse sparsity patterns.

In this work, we introduce \textbf{RSH-SpMM}, a row-structured hybrid GPU SpMM framework built on the \emph{RS-Tile} representation, a locality-aware reordering pass, and adaptive TC/CUDA partitioning. RS-Tile produces dense-aligned fragments for Tensor Cores while isolating structurally incompatible rows into a minimal-overhead CUDA path, enabling fine-grained, data-dependent execution.

Our contributions are summarized as follows:
\begin{itemize}
    \item We present \textbf{RS-Tile}, a compact row-structured representation that exposes Tensor-Core-aligned fragments while routing irregular rows to a CUDA execution path, reducing metadata overhead and improving tile density.

    \item We develop a \textbf{fine-grained hybrid execution strategy} that integrates adaptive row partitioning, a load-balanced pipelined Tensor-Core kernel, and a minimal-overhead CUDA kernel. This design avoids rigid tiling rules and eliminates the need for maintaining dual sparse indexing structures.

    \item We propose a \textbf{locality-aware reordering} technique that groups structurally similar rows to reduce tile fragmentation and improve MMA utilization.

    \item Using diverse real-world workloads, we evaluate RSH-SpMM against state-of-the-art CUDA-core, Tensor-Core, and hybrid GPU SpMM baselines. RSH-SpMM achieves \textbf{1.27$\times$--6.13$\times$} speedups, demonstrating stable performance gains across a wide variety of sparsity distributions.
\end{itemize}

\section{Background and Motivation}

\subsection{Modern GPU Architecture}

Modern GPUs adopt a massively parallel design centered on Streaming Multiprocessors (SMs), each integrating CUDA cores, Tensor Cores, warp schedulers, and a unified L1/shared-memory subsystem. Under the SIMT execution model~\cite{nvidia_cuda_c_2024}, threads operate in 32-lane warps, enabling efficient fine-grained parallelism for regular data-parallel workloads.

CUDA cores are general-purpose SIMT units that, despite lacking explicit support for irregular memory access, provide finer-grained control than Tensor Cores for unstructured work. Tensor Cores, first introduced in Volta and further enhanced in subsequent GPU generations~\cite{nvidia_volta_2017,nvidia_ampere_2020,nvidia_hopper_2022}, perform warp-level matrix-multiply-accumulate (MMA) operations on fixed-size tiles and deliver substantially higher throughput when data can be prepared as dense, aligned fragments. The surrounding memory hierarchy, combining high-bandwidth global memory, a unified L2 cache, and low-latency L1/shared memory, facilitates efficient tile buffering and reuse. General matrix-matrix multiplication (GEMM) workloads align well with this architecture: Tensor Cores execute most of the compute-intensive work, while CUDA cores handle auxiliary control and memory operations. This complementary capability motivates hybrid execution strategies that assign structured regions of sparse workloads to Tensor Cores while routing irregular portions to CUDA cores.

\subsection{SpMM on GPUs}

Sparse Matrix-Matrix Multiplication (SpMM) computes $C = A \times B$, where $A$ is sparse and $B$ is dense. It is a fundamental kernel in Graph Neural Networks (GNNs), sparse deep learning, graph analytics, and scientific computing. 
In contrast to dense GEMM, whose inputs can be tiled into perfectly aligned blocks for Tensor Core execution, real-world sparse matrices exhibit strong structural irregularity that fundamentally misaligns SpMM with the execution patterns of modern GPUs.

\begin{figure}[h]
  \centering
  \includegraphics[width=\linewidth]{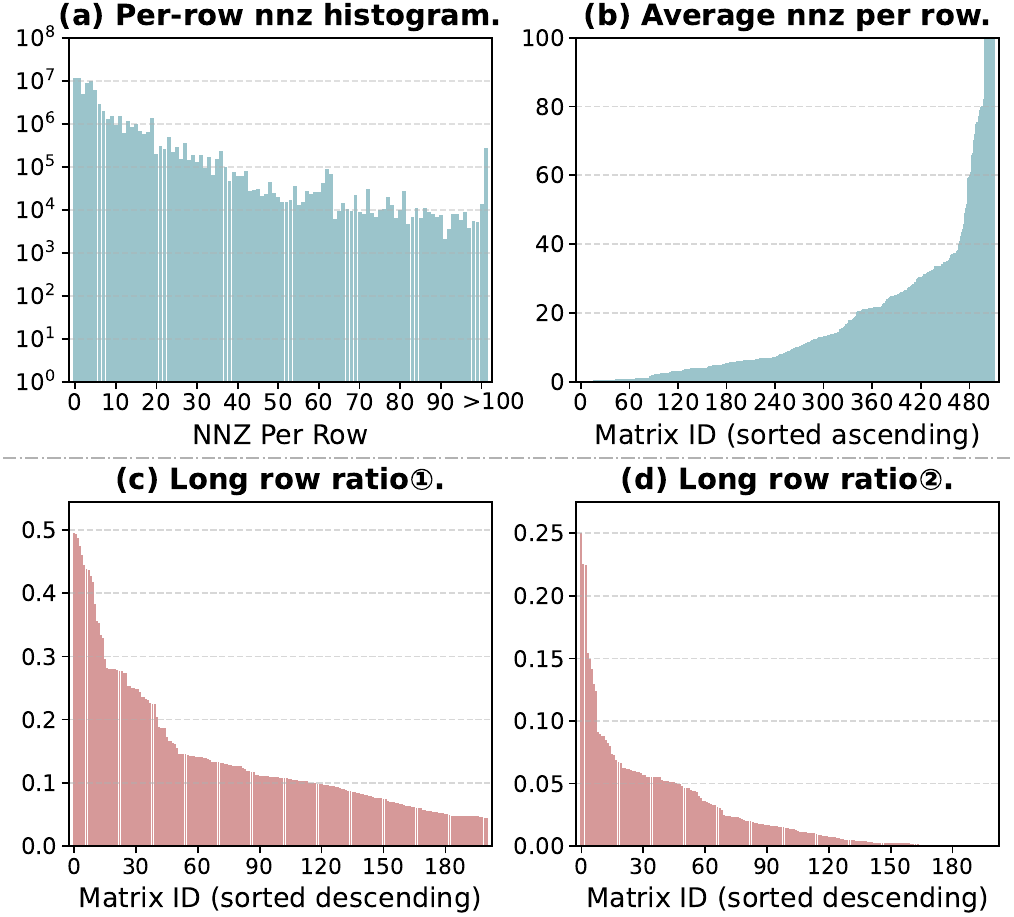}
  \caption{Structural characteristics of matrices in the SuiteSparse collection.
(a) Distribution of row counts grouped by the number of nonzeros (nnz).
The long-row ratios~\ding{172} and~\ding{173} correspond to rows with
$\mathrm{nnz} > 2\times nnz_{\text{mean}}$ and $\mathrm{nnz} > 4\times nnz_{\text{mean}}$, respectively. }
   \label{fig:matrixnnzinfo}
\end{figure}

Figure~\ref{fig:matrixnnzinfo} summarizes our empirical analysis of per-row nonzeros (nnz) statistics across the SuiteSparse Matrix Collection~\cite{suitesparse}.  
We highlight three key observations:

\begin{itemize}
    \item \textbf{Heavy-tailed row-wise nnz distribution.}  
    Most rows contain only a few nonzeros, whereas a small subset has nnz counts several orders of magnitude higher, reflecting pronounced structural imbalance across rows.

    \item \textbf{Substantial cross-matrix variation in mean nnz.}  
    The average nnz per row differs widely across matrices, reflecting diverse global sparsity structures that cannot be handled effectively by a single fixed execution strategy.

    \item \textbf{Non-negligible fraction of structurally ``long'' rows.}  
    Many matrices contain rows whose nnz counts exceed $2\text{--}4\times$ the matrix average, indicating pronounced row-level imbalance even among neighboring rows within the same matrix.
\end{itemize}

These observations imply that both inter-matrix and intra-matrix sparsity patterns vary rapidly and unpredictably. As a consequence, GPU execution behavior becomes highly sensitive to local structural properties. Many regions naturally fail to expose the dense substructures needed by Tensor Cores; at the same time, the irregular nnz distribution disrupts coalesced memory accesses, induces load imbalance, and lowers utilization of SIMT execution. Traditional formats such as Compressed Sparse Row (CSR)~\cite{csr} and Coordinate (COO)~\cite{coo} therefore struggle to deliver high and stable performance on modern GPUs.

\subsection{Motivation}

Although recent advances in CUDA-core kernels, Tensor-Core tiling, and hybrid execution have improved GPU SpMM performance, profiling across representative baselines reveals that existing approaches remain fundamentally misaligned with the fine-grained structural heterogeneity of real sparse matrices.

Tensor-Core approaches such as TC-GNN~\cite{tc-gnn}, DTC-SpMM~\cite{dtc-spmm}, and Acc-SpMM~\cite{acc-spmm} rely on fixed row windows and rigid MMA tile geometries to form pseudo-dense fragments. Real matrices, however, contain many short or isolated rows and exhibit abrupt density drops, causing the effective nonzero ratio of $16{\times}8$ or $8{\times}8$ tiles to fall to \textbf{14.5\%--20.4\%} on typical graphs and as low as \textbf{6.3\%--11.8\%} on highly irregular ones. This sharply lowers active MMA utilization; Acc-SpMM sustains only \textbf{5.6\%} useful MMA cycles on average, with most cycles stalled on padding and sparse fragments. Balanced variants attempt to compensate but introduce row remapping, extra accumulation, and heavy atomic operations, often running up to \textbf{1.4$\times$} slower than versions without atomics. These issues arise from the intrinsic inflexibility of fixed window sizes and tile shapes, which cannot adapt to rapid row-level sparsity variation.

Hybrid CUDA-Tensor-Core schemes such as HC-SpMM~\cite{hc-spmm} and HR-SpMM~\cite{hr-spmm} attempt to combine compute units by routing rows to different execution paths, typically using global sparsity statistics or simple per-row thresholds. Such coarse-grained heuristics ignore key structural cues such as inter-row similarity, column-span overlap within windows, and local density patterns that determine whether high-quality Tensor-Core tiles can form. Consequently, many rows assigned to the TC path still produce low-density fragments, while a large fraction of nonzeros (sometimes exceeding 50\%) is diverted to the CUDA path. This forces CUDA cores to process workloads poorly matched to their memory-bound execution model and leaves Tensor Cores underutilized, resulting in only modest gains over pure CUDA-core kernels.

These limitations reveal that existing SpMM designs lack a mechanism that adapts to sparsity where it actually varies and coordinates heterogeneous execution effectively. A high-performance GPU SpMM kernel should therefore satisfy three key requirements:
(i) adapt row-group decisions to local sparsity fluctuations and inter-row similarity, assigning work to Tensor Cores only when high-quality tiles can form while routing short or structurally volatile rows to a locality-preserving CUDA path;
(ii) use representations and execution paths suited to the strengths of each compute unit, enabling Tensor Cores to operate on dense, MMA-friendly tiles while keeping the CUDA path lightweight and structurally coherent; and
(iii) sustain pipeline stability through adaptive load balancing that keeps SMs well utilized while minimizing atomic synchronization.

These requirements motivate \textbf{RSH-SpMM}, a row-structured, fine-grained hybrid SpMM framework that aligns sparse-matrix structure with the heterogeneous capabilities of modern GPUs.

\section{Design of RSH-SpMM}

\subsection{Overview}

\begin{figure}[h]
  \centering
  \includegraphics[width=\linewidth]{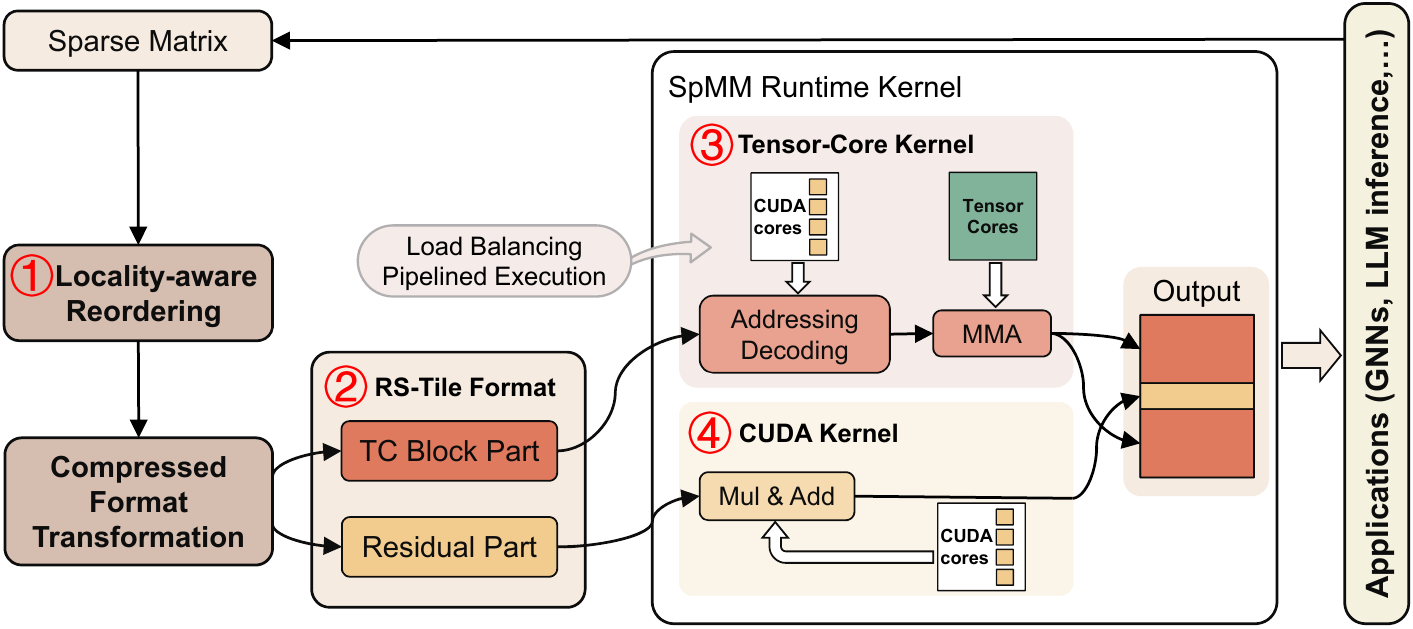}
  \caption{Overview of RSH-SpMM. }
   \label{fig:overview}
\end{figure}

As shown in Figure~\ref{fig:overview}, RSH-SpMM is organized as a hybrid execution workflow with a row-structured design that aligns sparse-matrix structure with the heterogeneous capabilities of modern GPUs. \redcirc{1} \emph{Locality-aware reordering} is applied to enhance structural similarity among neighboring rows, improving local density and facilitating the formation of high-quality tiles. The reordered matrix is then converted into \redcirc{2} \emph{RS-Tile} representation, which aggregates consecutive rows into the Tensor-Core-friendly part while directing short or isolated rows into the CUDA residual part. This organization preserves tile regularity and prevents low-density rows from undermining Tensor Core utilization. SpMM computation is executed through two specialized kernels: \redcirc{3} \emph{Tensor-Core Kernel}, which uses CUDA cores for decoding and address reconstruction before issuing MMA operations on Tensor Cores, and \redcirc{4} \emph{CUDA Kernel}, which processes the residual rows through a lightweight multiply-add path optimized for fine-grained irregularity, with the two kernels handling disjoint portions of the output.

These components form a unified hybrid execution framework: Tensor Cores handle locally regular regions for high throughput, while CUDA cores provide an efficient fallback for highly irregular structures. This coordinated design delivers stable performance across diverse sparsity patterns.

\subsection{Compressed Format: RS-Tile}

\begin{figure}[h]
  \centering
  \includegraphics[width=\linewidth]{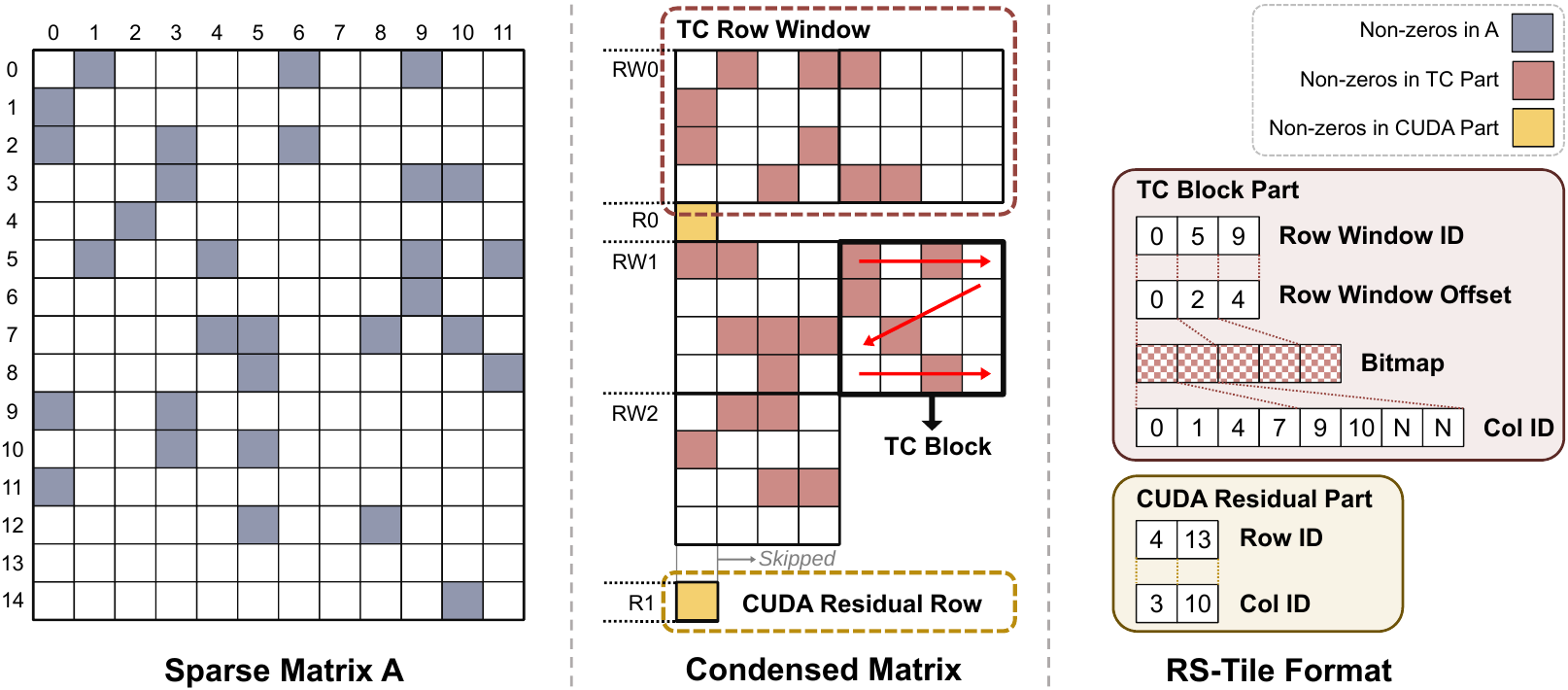}
  \caption{RS-Tile compressed format of sparse matrix A. A $4\times 4$ tile is shown for convenience in illustration, whereas our implementation employs the native $8\times 8$ MMA footprint on Tensor Cores.}
   \label{fig:compressed-format}
\end{figure}

We optimize the sparse matrix storage format by introducing a row-structured tiling scheme, referred to as \emph{RS-Tile}. As shown in Figure~\ref{fig:compressed-format}, the matrix is decomposed into two disjoint parts: a TC part and a CUDA residual part. Rows with very few nonzeros or column patterns that differ significantly from their neighbors would degrade tile density and inflate the number of tiles. Such rows are therefore separated into the CUDA part, while the remaining structurally coherent regions are grouped into TC row windows. Both parts use distinct indexing structures tailored to the characteristics of Tensor Cores and CUDA cores.

\textbf{TC Block Part.}
Each row window is compacted by collecting all distinct column indices accessed by its nonzeros and remapping them into a contiguous column space. The window is then partitioned into fixed-size $8\times 8$ TC blocks that match the MMA tile geometry (the figure uses a $4\times 4$ tile for illustration). The resulting TC block region is represented by four arrays: the \emph{Row Window ID}, which stores the global starting row of each window; the \emph{Row Window Offset}, which records the prefix-sum position of each TC block in the bitmap stream; the \emph{Bitmap}, a compact bitmask encoding nonzero locations within each block; and the \emph{Col ID}, which stores the original column indices corresponding to block nonzeros.
For the last block of a row window, if the available elements do not fully occupy an $8\times 8$ footprint, we insert lightweight dummy entries to preserve strict geometric alignment and avoid unnecessary memory divergence during fragment construction.

\textbf{CUDA Residual Part.}
Remaining isolated short rows are stored in a CUDA-core-friendly representation. Because these rows contain very few nonzeros, the CSR row-pointer array is omitted, and each row is kept in a compact small-row format consisting only of its \emph{Row ID} (the global row index) and its \emph{Col ID} entries (the original column indices of its nonzeros). Empty rows are skipped entirely to avoid redundant indexing and reduce processing overhead.

After compression, the sparse matrix is organized into:  
(i) a TC block part composed of bitmap-compressed, tile-aligned row windows, and  
(ii) a CUDA residual part storing short and structurally inconsistent rows. This 
division improves tile density, reduces window-level imbalance, and provides a regularized memory layout suitable for high-throughput Tensor Core execution.

\subsection{Balanced Fine-grained Kernel}

\subsubsection{Fine-Grained Row Partition}

\begin{algorithm}[t]
\caption{Row-Level Execution Partitioning}
\label{alg:partition}
\begin{algorithmic}[1]

\State \textbf{Input:} CSR matrix $A$, window size $W$
\State \textbf{Output:} TC windows $\mathcal{W}$, CUDA residual rows $\mathcal{C}$

\State $n_{\text{row}} \gets \text{num\_rows}(A)$
\State $N_{\text{nz}} \gets \textsc{NNZ}(A)$
\State $(\tau_{\text{nnz}}, \tau_{\text{inc}}) \gets \textsc{EstimateThreshold}(n_{\text{row}}, N_{\text{nz}})$
\Statex

\State $\mathcal{W} \gets \emptyset$, $\mathcal{C} \gets \emptyset$, $r \gets 0$

\While{$r < n_{\text{row}}$}
    \State $nz_r \gets \textsc{NNZ}(A[r])$
    \If{$nz_r = 0$}
        \State $r \gets r + 1$
        \State \textbf{continue}
    \EndIf
    \State $U_{\text{full}} \gets$ distinct columns in rows $[r, r+W-1]$
    \State $U_{\text{excl}} \gets$ distinct columns in rows $[r+1, r+W-1]$
    \State $\Delta \gets |U_{\text{full}}| - |U_{\text{excl}}|$
    \Statex
    \If{$\Delta < \tau_{\text{inc}}$ \textbf{and} $nz_r \le \tau_{\text{nnz}}$}
        \LCommenttab{low-impact row, assign to CUDA path}
        \State $\mathcal{C} \gets \mathcal{C} \cup \{r\}$ 
        \State $r \gets r + 1$
    \Else
        \State append window $[r, r+W-1]$ to $\mathcal{W}$
        \State $r \gets r + W$
    \EndIf
\EndWhile

\State \Return $(\mathcal{W}, \mathcal{C})$

\end{algorithmic}
\end{algorithm}

RSH-SpMM performs a lightweight row-level assignment before constructing row windows to identify short or structurally incompatible rows that cannot be effectively merged into Tensor-Core windows. Such rows are removed from the main Tensor-Core execution path and delegated to the CUDA residual path, which is better suited for handling fine-grained sparsity irregularity.

As shown in Algorithm~\ref{alg:partition}, the system sequentially scans all rows and makes decisions based on two key factors. First, rows with a large number of nonzeros exhibit higher computational intensity and typically contribute substantial column coverage; these rows are therefore considered direct Tensor-Core candidates. Second, for rows with fewer nonzeros and lower arithmetic intensity, we estimate their local structural impact by comparing the set of distinct columns in a candidate window that includes the current row with the corresponding window that excludes it. If the increment in covered columns is negligible, the row provides little useful structural contribution and may even degrade the density of the window; such low-impact rows are therefore assigned to the CUDA residual path. Conversely, rows that remain structurally consistent with their neighbors and do not significantly increase tile formation cost are merged into the appropriate row window and processed by the Tensor-Core execution path.

Importantly, dispatching short rows to CUDA cores is motivated not only by tile-density considerations. CUDA cores inherently handle lightweight operations more efficiently than Tensor Cores: their execution path is simpler and does not require the construction of full MMA tiles or fragment states, resulting in lower register pressure and scheduling overhead. In contrast, short rows often fail to sustain adequate pipeline utilization in Tensor-Core kernels, as the cost of tile preparation and memory staging may outweigh the actual arithmetic work. Assigning these rows to CUDA cores therefore improves overall resource efficiency and enhances GPU concurrency and stability.

Through this fine-grained row-level decision mechanism, RSH-SpMM maintains compact, dense, and structurally coherent Tensor-Core windows while routing highly irregular or isolated rows to a more appropriate execution path. This row-centric hybrid design mitigates performance penalties caused by sparsity fragmentation, improves load balance within windows, and maximizes overall throughput on modern GPUs.

\subsubsection{Pipelined Kernel Execution}

\begin{figure}[h]
  \centering
  \includegraphics[width=\linewidth]{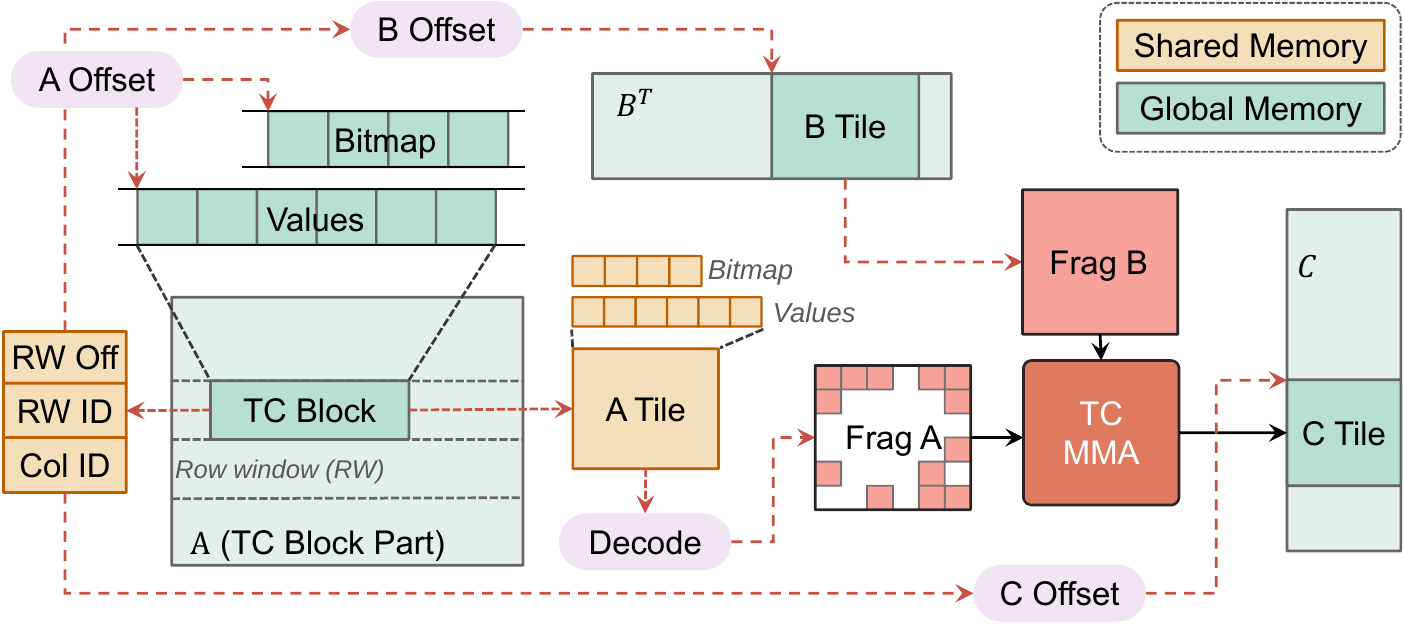}
  \caption{Dataflow of the Tensor Core computation. }
   \label{fig:datamovement}
\end{figure}

The RS-Tile format enables a highly structured execution flow for the Tensor-Core (TC) part. Our TC kernel is organized around four key components: (i) a global--to--shared--to--register dataflow tailored for sparse tiles, (ii) a bitmap-based shared-memory decoding mechanism, (iii) a two-stage double-buffer pipeline that overlaps memory movement and MMA computation, and (iv) a register-level fragment assembly scheme that maximizes Tensor Core utilization.

Figure~\ref{fig:datamovement} shows the data movement pattern of the TC kernel. Sparse tiles of $A$ are first prefetched from global memory into shared memory, where values and bitmap segments are staged together. During the shared-memory--to--register transfer, the bitmap guides the expansion of nonzeros into a dense $(8\times 8)$ MMA footprint. We employ warp-level bit operations to determine each active element’s position in the decoded fragment, with \texttt{\_\_popcll()} computing prefix counts within the bitmap. This decoding procedure eliminates pointer chasing and avoids additional shared-memory accesses, producing a fully populated MMA operand directly in registers.

\begin{algorithm}[t]
\caption{Pipelined Tensor-Core Kernel in RSH-SpMM}
\label{alg:tc-pipeline}
\begin{algorithmic}[1]

\State \textbf{Input:} RS-Tile format $A$, dense matrix $B$
\State \textbf{Output:} partial results in $C$
\Statex

\State $start\_blk \gets \texttt{RowWindowOffset}[\texttt{bid.x}]$
\State $end\_blk \gets \texttt{RowWindowOffset}[\texttt{bid.x}+1]$
\LComment{Double buffers in shared memory for A \& in registers for B}
\State \textbf{shared} \texttt{A\_value[2]}, \texttt{A\_ColID[2]}
\State \textbf{register} \texttt{fragA}, \texttt{fragB[2]}, \texttt{fragC} $\gets 0$

\Statex
\State \textbf{Prefetch phase}
\State \texttt{A\_value[0/1]} $\gets \texttt{global\_A\_value[start\_blk/start\_blk+1]}$
\State \texttt{A\_ColID[0/1]} $\gets \texttt{global\_A\_ColID[start\_blk/start\_blk+1]}$

\State \texttt{sync}()
\State \texttt{fragB[0]} $\gets$ load $8 \times 16$ tile from B$[start\_blk]$

\Statex
\State \textbf{Main phase}
\For{$t \gets start\_blk$ \textbf{to} $end\_blk - 1$}
    \State $\texttt{rdPart} \gets (t - start\_blk) \bmod 2$
    \State $\texttt{stPart} \gets 1 - \texttt{rdPart}$
    \If{$t + 1 < end\_blk$}
        \LCommenttabtab{Prefetch next\_blk of $A$ and $B$ tile into the other buffer}
        \State \texttt{A\_value[stPart]} $\gets \texttt{Async.load next\_blk}$ \Comment{value}
        \State \texttt{A\_ColID[stPart]} $\gets \texttt{Async.load next\_blk}$ \Comment{ColID}
        \State \texttt{fragB[stPart]} $\gets$ load $8 \times 16$ tile from \texttt{B}$[next\_blk]$
    \EndIf
    \LCommenttab{Decode current sparse values to register}
    \State \texttt{bitmap} $\gets \texttt{global\_bitmap[}t\texttt{]}$
    \State \texttt{fragA} $\gets \texttt{DecodeTile}(\texttt{A\_value[rdPart]}, \texttt{bitmap})$
    \LCommenttab{Tensor Core MMA on reconstructed fragments}
    \State \texttt{fragC} $\gets \texttt{mma.sync.m16n8k8}(\texttt{fragB[rdPart]}, \texttt{fragA}, \texttt{fragC})$
    \State \texttt{wait\_group}()
    \State \texttt{sync}()
\EndFor
\Statex
\State \textbf{Drain phase}
\State \texttt{OutBase} $\gets \texttt{A\_RowWindowID[bid.x]}$
\State \texttt{C[OutBase]} $\gets \texttt{fragC} $

\end{algorithmic}
\end{algorithm}

\begin{figure}[h]
  \centering
  \includegraphics[width=0.9\linewidth]{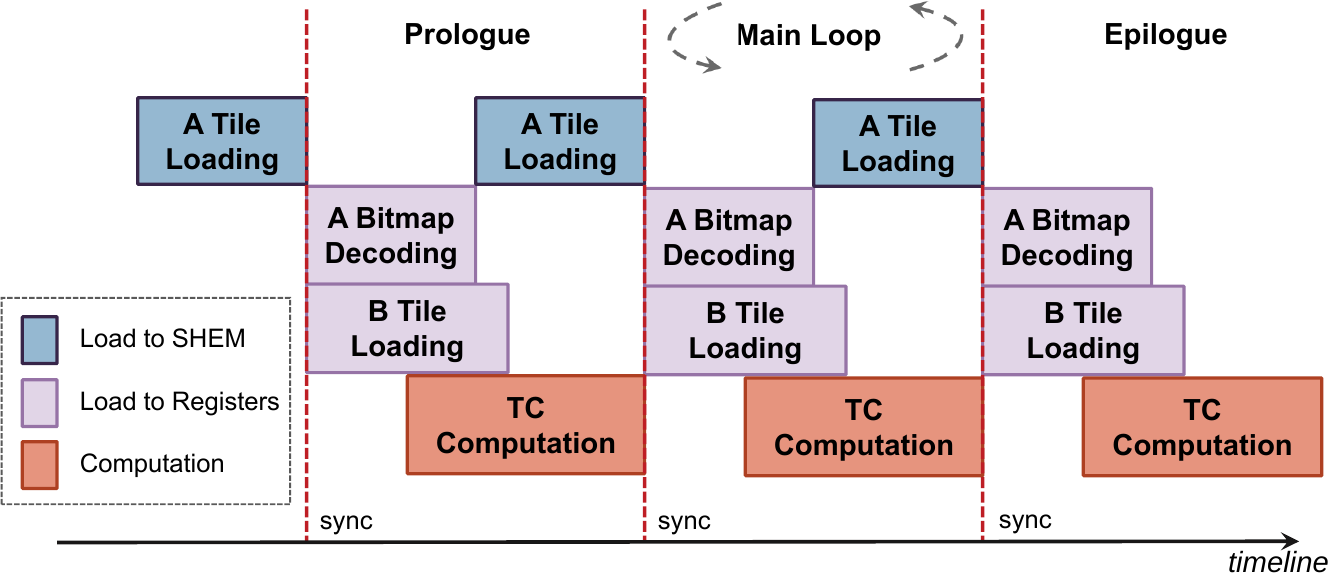}
  \caption{Execution pipeline of the Tensor-Core kernel.}
   \label{fig:pipeline}
\end{figure}

To maintain continuous Tensor Core utilization, we adopt a two-stage double buffer for both sparse tiles of $A$ and dense tiles of $B$. One buffer is consumed by the MMA instruction in the current iteration, while the other buffer is asynchronously filled using \texttt{cp.async}. As illustrated in Figure~\ref{fig:pipeline} and detailed in Algorithm~\ref{alg:tc-pipeline}, each iteration overlaps four operations:
\begin{itemize}
    \item prefetching the next sparse tile of $A$ into shared memory;
    \item loading the next $(8\times16)$ tile of $B$ directly into registers;
    \item bitmap-guided decoding of the active sparse tile into \texttt{fragA};
    \item issuing Tensor Core MMA on (\texttt{fragA}, \texttt{fragB}).
\end{itemize}

Once the pipeline prologue completes, the kernel enters a steady-state pipeline phase in which global-memory transfers, shared-memory staging, register-level decoding, and MMA execution proceed in parallel under an alternating double-buffer schedule. This design hides global-memory latency and ensures Tensor Cores are consistently fed with ready-to-use fragments even under highly irregular sparsity patterns.

\begin{figure}[h]
  \centering
  \includegraphics[width=0.95\linewidth]{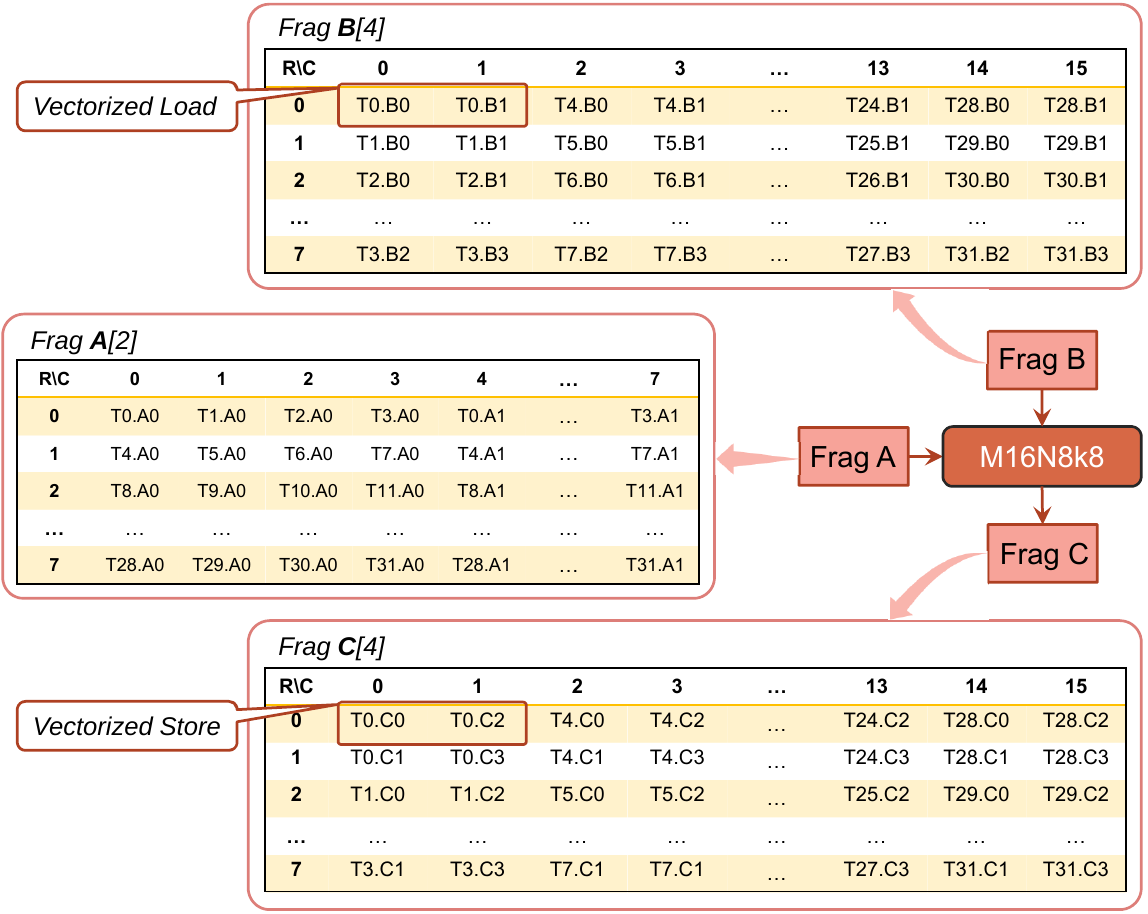}
  \caption{Register mapping of MMA operand fragments for Tensor Core execution. }
   \label{fig:register}
\end{figure}

As shown in Figure~\ref{fig:register}, for each thread, each $(8\times 8)$ tile of $A$ is expanded into two subfragments (\textbf{A[2]}), while each $(8\times 16)$ tile of $B$ is divided into four register fragments (\textbf{B[4]}). A lightweight in-register transpose converts the naturally loaded $(8\times 16)$ layout of $B$ into the $(16\times 8)$ operand format required by \texttt{mma.sync.m16n8k8}. We redesign the register mapping of $B$ such that, for each thread, every two $B$ register fragments are placed contiguously, enabling them to be fetched through one vectorized load, and the corresponding $C$ fragments are similarly aligned for vectorized stores.

\begin{figure}[h]
  \centering
  \includegraphics[width=0.9\linewidth]{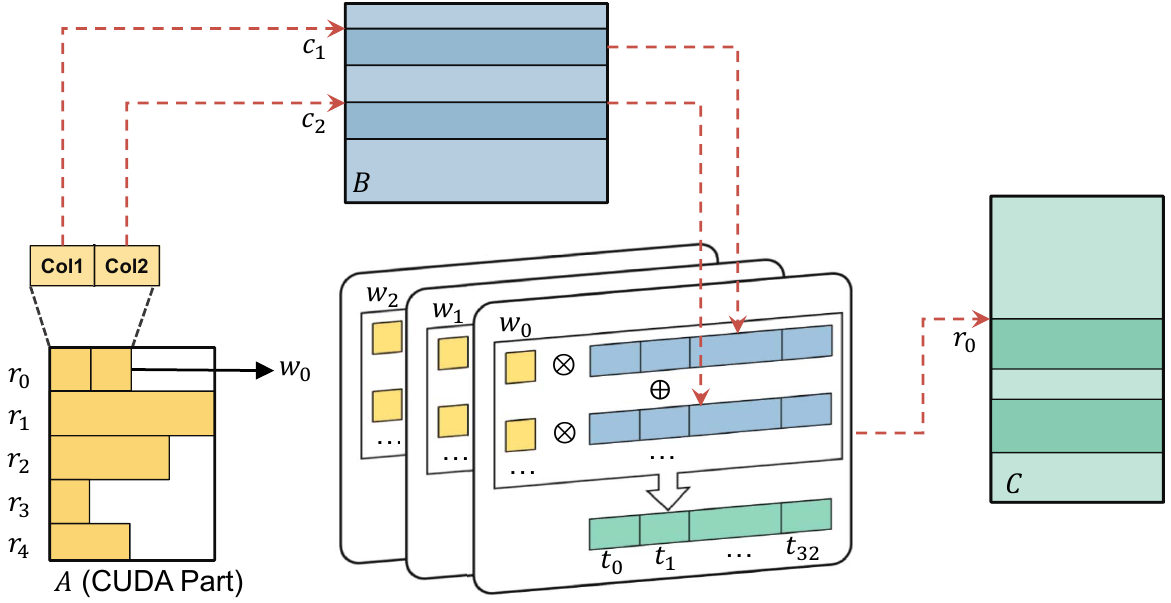}
  \caption{CUDA execution path for residual rows.}
   \label{fig:cudapart}
\end{figure}

While the Tensor-Core pipeline dominates the execution of dense-aligned windows, the residual rows assigned to CUDA cores follow a low-overhead execution path. As shown in Figure~\ref{fig:cudapart}, these rows are processed using a simple fused-load-compute kernel without tile reconstruction or pipeline staging. Because these rows contain only a few nonzeros, the CUDA path avoids the overhead associated with forming Tensor-Core tiles and achieves higher efficiency through reduced register pressure, minimal synchronization, and a short critical path. This complementary fast path ensures that both high-density and low-intensity regions are executed with appropriate compute units, maintaining overall throughput stability.

\subsubsection{Adaptive Load Balance}

\begin{figure}[t]
    \centering

    \begin{subfigure}{0.85\linewidth}
        \centering
        \includegraphics[width=\linewidth]{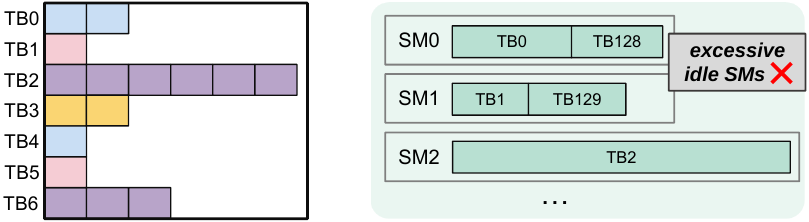}
        \caption{No balancing for TC-GNN. }
        \label{fig:lb_tcg}
    \end{subfigure}

    \begin{subfigure}{0.85\linewidth}
        \centering
        \includegraphics[width=\linewidth]{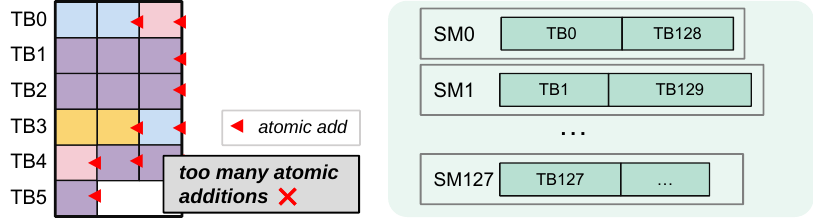}
        \caption{Fixed balancing for DTC-SpMM and Acc-SpMM. }
        \label{fig:lb_dtc}
    \end{subfigure}

    \begin{subfigure}{0.87\linewidth}
        \centering
        \includegraphics[width=\linewidth]{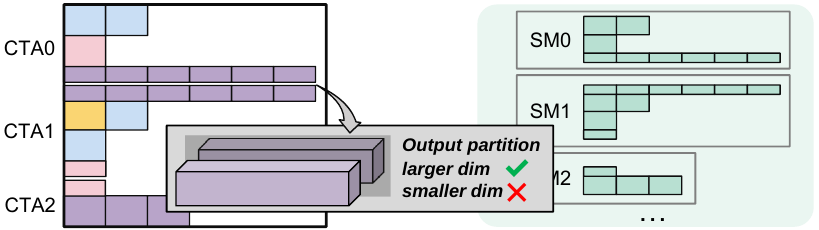}
        \caption{Output balancing for Voltrix.}
        \label{fig:lb_vol}
    \end{subfigure}

    \begin{subfigure}{0.87\linewidth}
        \centering
        \includegraphics[width=\linewidth]{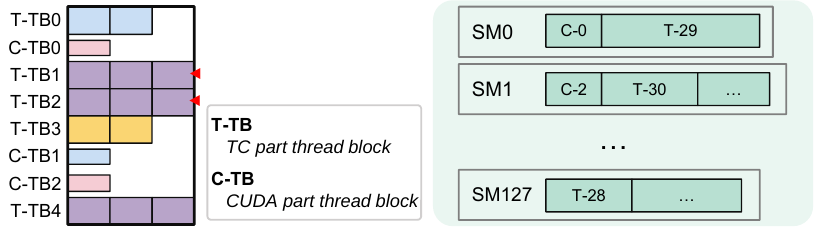}
        \caption{Adaptive load balancing for RSH-SpMM. }
        \label{fig:lb_ours}
    \end{subfigure}
    \caption{Comparison of load balancing strategies for SpMM kernels. TB denotes a thread block and CTA denotes a cooperative thread array.}
    \label{fig:lb_compare}
\end{figure}

Sparse matrices exhibit drastic variation in row density, and without proper balancing a Tensor-Core kernel may encounter poorly sized windows, uneven warp workloads, or cases where a single long row dominates the latency of the entire window. Thus, load balancing is essential for Tensor-Core-accelerated SpMM. As illustrated in Figure~\ref{fig:lb_compare}, existing Tensor-Core-accelerated methods generally adopt rigid or all-or-nothing strategies. TC-GNN (Figure~\ref{fig:lb_tcg}) separates rows by degree but does not address intra-window variation; DTC-SpMM and Acc-SpMM (Figure~\ref{fig:lb_dtc}) rely on fixed window sizes and cannot adapt to density fluctuations within the window; Voltrix (Figure~\ref{fig:lb_vol}) uses Cooperative Thread Array (CTA)-level partitioning to split work along the output dimension and avoid atomics. Operating at CTA granularity increases the complexity of managing irregular workloads, and the 16-column alignment requirement further limits load-balancing flexibility when feature sizes are small.

In contrast, RSH-SpMM (Figure~\ref{fig:lb_ours}) adopts a structure-driven and flexible balancing scheme. ``Super-long'' rows are selectively split based on matrix size, sparsity, and tile expansion behavior, whereas medium and short rows remain aggregated, avoiding unnecessary atomics or index rewriting for the majority of the workload. Extremely short or structurally incompatible rows are isolated into the CUDA residual path earlier, reducing load variance at the source and improving Tensor Core occupancy. This context-aware design adapts balancing granularity to actual sparsity patterns, neither ignoring imbalance nor enforcing rigid uniformity, yielding stable warp/block-level load balance and sustained Tensor Core throughput across highly heterogeneous matrices.

\subsection{Locality-aware Reordering}

To enhance the structural coherence of row windows and increase the likelihood that contiguous rows admit compact TC tiles under RS-Tile encoding, we introduce a locality-aware reordering stage prior to format construction. Each row $r$ is represented by its column-support set $S_r\subseteq\{1,\dots,n\}$ indicating the presence of nonzeros. To obtain a similarity measure that more directly reflects how two rows contribute to reducing adjacent structural variation, we employ a weighted Jaccard similarity (w-Jaccard). Let $d_j$ denote the global degree (column frequency) of column~$j$, where
\[
w_j = d_j^{-\alpha}, \qquad \alpha>0,
\]
which assigns lower influence to ubiquitous column features. The affinity between rows $r$ and $u$ is then defined as
\[
\text{sim}(r,u)
=
\frac{\sum_{j\in S_r\cap S_u} w_j}{\sum_{j\in S_r\cup S_u} w_j},
\]
which emphasizes agreements on informative low-frequency columns while diminishing the effect of high-frequency ones.

The reordering problem is formulated as minimizing the cumulative dissimilarity along the final row sequence:
\[
\min_{\pi}\;\sum_{i=1}^{m-1}\big(1-\text{sim}(\pi_i,\pi_{i+1})\big),
\]
a formulation that corresponds to finding a minimum-cost Hamiltonian path when distances are measured using the weighted Jaccard similarity metric.

\begin{algorithm}[t]
\caption{Locality-aware Reordering}
\label{alg:reorder}
\begin{algorithmic}[1]

\State \textbf{Input:} sparse matrix $A$
\State \textbf{Output:} reordered matrix $A_{\text{reordered}}$
\Statex

\LCommentbf{Step I: Candidate Construction}
\State $R \gets$ row of $A$
\State $F \gets$ representative features extracted from $A$
\State build inverted index on $F$
\For{$r \in R$}  
    \State $C(r) \gets$ rows sharing features with $r$
\EndFor

\Statex
\LCommentbf{Step II: $k$NN Graph Construction}
\For{$r \in R$}
    \State $\text{sim}(r,u)$ for $u \in C(r)$\Comment{compute \textit{w-Jaccard similarity} }
    \State $N_k(r) \gets$ top-$k$ neighbors of $r$ by $\text{sim}$
\EndFor
\State build undirected $k$NN graph $G$ from $\{N_k(r)\}$

\Statex
\LCommentbf{Step III: MST-based Global Ordering}
\State assign edge weights in $G$ using $\text{sim}$
\State $T \gets$ minimum spanning tree of $G$
\State $\pi \gets$ DFS traversal order on $T$

\Statex
\LCommentbf{Step IV: Local Refinement \& Isolation Adjustment}
\For{each window $W$ in $\pi$}
    \State apply 2-opt swaps in $W$ if adjacent cost decreases
\EndFor
\State $S \gets$ rows in $\pi$ that are locally isolated
\For{$r \in S$}
    \State $u \gets$ most similar row to $r$ in $R \setminus S$
    \State place $r$ adjacent to $u$ (or append to tail if none)
\EndFor

\Statex
\State \Return reordered matrix $A[\pi,:]$

\end{algorithmic}
\end{algorithm}

Since computing the optimal permutation over all $m$ rows requires exploring $O(m!)$ possible orderings, a search space that is computationally intractable for large matrices, we approximate the objective through the multi-stage procedure summarized in Algorithm~\ref{alg:reorder}. The first stage constructs candidate neighborhoods using an inverted index over column features, limiting similarity evaluation to structurally relevant row pairs and avoiding the cost of all-pairs comparisons. A $k$-nearest-neighbor (kNN) graph is constructed to capture the dominant local relationships in the sparsity pattern. From this graph, a minimum-spanning tree (MST) is extracted to impose a coherent global structure while preserving connectivity across neighborhoods. A depth-first traversal of the MST provides an initial permutation that retains the major affinity trends among rows. The final refinement stage improves local alignment: within each prospective row window, 2-opt exchanges are applied when beneficial to reduce adjacent dissimilarity. Rows that remain structurally isolated, with no meaningful affinity to surrounding neighbors, are placed near their closest matches to prevent breaks in locality.

This reordering yields a row sequence with consistently lower adjacent dissimilarity under the weighted Jaccard metric, producing windows with more compact aggregated column sets and fewer TC tiles under RS-Tile construction. Rather than altering the sparsity pattern, the method simply restructures row order to reveal the latent structural continuity already present in the matrix.

\section{Evaluation}

\begin{figure*}[t]
  \centering
  \includegraphics[width=\textwidth]{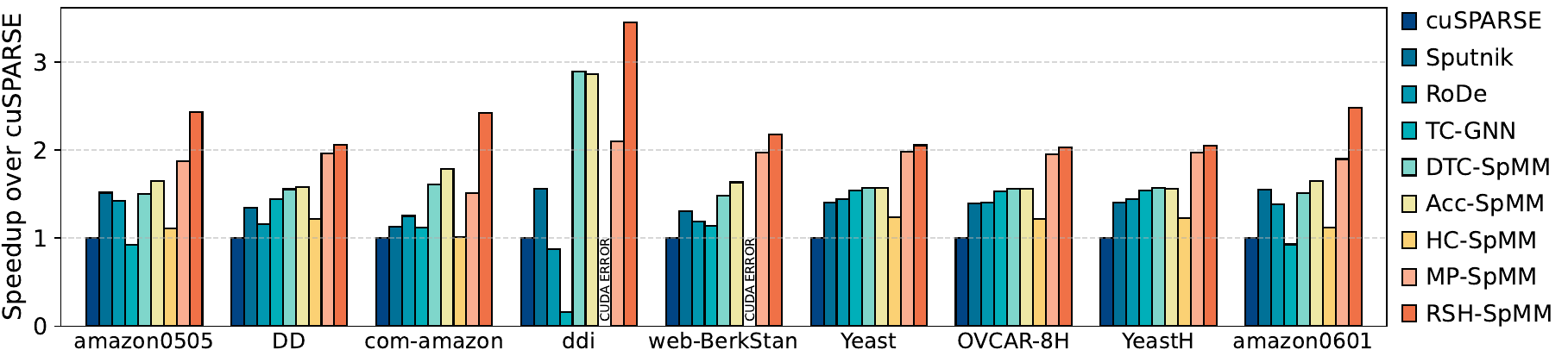}
  \caption{Kernel-level SpMM throughput on 9 representative real-world matrices on RTX~4090, reported as normalized speedup over cuSPARSE. }
   \label{fig:result}
\end{figure*}

\begin{figure*}[t]
  \centering
  \includegraphics[width=\textwidth]{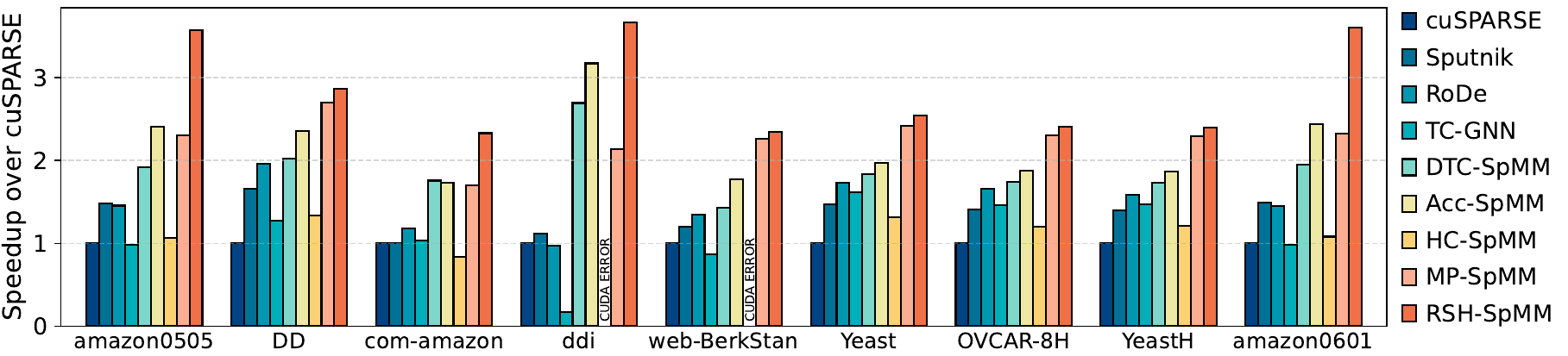}
  \caption{Kernel-level SpMM throughput on 9 representative real-world matrices on RTX~3090, reported as normalized speedup over cuSPARSE. }
   \label{fig:result_3090}
\end{figure*}

\subsection{Experimental Setup}

\noindent\textbf{Hardware Platform.}
We evaluate our implementation on two NVIDIA GPUs with distinct architectures: RTX~4090 (Ada Lovelace, compute capability 8.9, 24\,GB memory) and RTX~3090 (Ampere, compute capability 8.6, 24\,GB memory). Reported results are taken from warmed-up kernel executions.

\noindent\textbf{Datasets.}
We evaluate RSH-SpMM on 9 representative real-world matrices~\cite{snap,tc-gnn,dgl} commonly used in GNN and sparse learning workloads. These datasets cover a broad spectrum of sparsity patterns, including highly skewed degree distributions and heterogeneous local densities. Table~\ref{tab:dataset} summarizes their basic statistics.
To evaluate robustness under diverse sparsity patterns, we additionally include 512 matrices from the SuiteSparse Matrix Collection~\cite{suitesparse}, which is widely used for benchmarking sparse linear algebra kernels. We select matrices with at least 5K rows, 5K columns, and 100K nonzeros to ensure that the workloads are sufficiently large and structurally varied. Spanning a wide range of application domains and sparsity characteristics, the SuiteSparse collection serves as a standard benchmark for assessing SpMM robustness and generality.

\begin{table}[t]
\centering
\caption{Statistics of representative real-world matrices used for evaluation.
Superscripts indicate data sources:
$^{1}$TC-GNN~\cite{tc-gnn},
$^{2}$SNAP~\cite{snap},
$^{3}$DGL~\cite{dgl}.}
\label{tab:dataset}
\begin{tabular}{lrrr}
\toprule
Dataset & Nodes & Row\&Col & COO Size \\
\midrule
com-amazon$^{2}$     & 548{,}551     &   925{,}872     & 13.8 MB \\
ddi$^{3}$            &   4{,}267     & 2{,}140{,}089   & 23.1 MB \\
DD$^{1}$             &   334{,}925   & 1{,}686{,}092   & 24.6 MB \\
amazon0505$^{2}$     & 410{,}236     & 3{,}356{,}824   & 48.5 MB \\
amazon0601$^{2}$     & 403{,}394     & 3{,}387{,}388   & 48.9 MB \\
Yeast$^{1}$          & 1{,}710{,}902 & 3{,}636{,}546   & 57.9 MB \\
OVCAR-8H$^{1}$       & 1{,}889{,}542 & 3{,}946{,}402   & 63.3 MB \\
YeastH$^{1}$         & 3{,}138{,}114 & 6{,}487{,}230   & 107.0 MB \\
web-BerkStan$^{2}$   & 685{,}230     & 7{,}600{,}595   & 112.3 MB \\
\bottomrule
\end{tabular}
\end{table}

\noindent\textbf{Baselines.}
We compare RSH-SpMM with both CUDA-core and Tensor-Core-accelerated SpMM implementations. 
For CUDA-core baselines, we include the vendor CSR kernel \textbf{cuSPARSE}~\cite{nvidia_cusparse_2024}, \textbf{Sputnik}~\cite{sputnik}, a warp-cooperative sparse kernel, and \textbf{RoDe}~\cite{rode}, a row-distribution-based SpMM kernel optimized for load balance and memory efficiency.
For Tensor-Core-accelerated baselines, we evaluate \textbf{TC-GNN}~\cite{tc-gnn}, which maps GNN-style SpMM to Tensor Cores, \textbf{DTC-SpMM}~\cite{dtc-spmm}, which constructs row-window-based TC tiles, and \textbf{Acc-SpMM}~\cite{acc-spmm}, which employs a bitmap-oriented TC format for $8{\times} 8$ tiles. 
We further include hybrid approaches and methods that target Sparse Tensor Cores (SpTCs): \textbf{HC-SpMM}~\cite{hc-spmm}, which partitions structured and irregular regions across CUDA cores and Tensor Cores, and \textbf{MP-SpMM}~\cite{mp-spmm}, which transforms unstructured sparsity into 2:4/1:2 structured patterns via a matching-and-padding scheme to leverage Sparse Tensor Cores. 

\noindent\textbf{Problem Configuration.}
All SpMM evaluations compute $C = A \times B$ with a dense feature matrix $B \in \mathbb{R}^{n \times d}$. We benchmark four commonly used feature dimensions, $d \in \{64, 128, 256, 512\}$, which reflect typical settings in GNNs and sparse learning workloads. During Tensor Core execution, FP16 inputs are multiplied using the Tensor Core MMA pipeline, while all partial sums are accumulated in FP32. The final output $C$ is stored in FP32 to ensure consistent numerical behavior across baselines.

\subsection{Overall Performance}
We evaluated the SpMM kernel performance of RSH-SpMM on 9 representative real-world matrices, with results averaged over feature dimensions $64$, $128$, $256$, and $512$. Figures~\ref{fig:result} and~\ref{fig:result_3090} report the normalized speedup over cuSPARSE on RTX~4090 and RTX~3090, respectively. Across all datasets, RSH-SpMM consistently achieves the highest kernel-level throughput among all evaluated baselines. Compared with cuSPARSE, it delivers an average speedup of \textbf{2.35$\times$} on RTX~4090 and \textbf{2.86$\times$} on RTX~3090.

Compared with CUDA-core implementations, including cuSPARSE, Sputnik, and RoDe, RSH-SpMM demonstrates clear and consistent gains. While methods such as Sputnik and RoDe can perform competitively on matrices with relatively regular sparsity patterns or high row densities, CUDA-core kernels are fundamentally limited in compute throughput compared with Tensor Cores, and their performance degrades on irregular workloads with long or imbalanced rows. By leveraging adaptive partitioning and hybrid execution, RSH-SpMM achieves average improvements of \textbf{2.06$\times$--2.61$\times$} over CUDA-core baselines.
Among Tensor-Core-based methods, RSH-SpMM shows consistent performance advantages. It outperforms TC-GNN by up to \textbf{6.13$\times$} on average over the reported cases, after excluding configurations where TC-GNN encounters CUDA runtime errors or exhibits prohibitively long execution times, reflecting the sensitivity of rigid tiling without explicit load balancing under uneven row densities. Compared with more optimized Tensor-Core designs such as DTC-SpMM and Acc-SpMM, which improve locality through reordering and window-based tiling, RSH-SpMM continues to achieve steady gains, as fixed windowing strategies can lead to underfilled MMA fragments when sparsity varies across rows.
Hybrid designs such as HC-SpMM rely on multi-path execution but are limited by coarse-grained partitioning. Excluding configurations that encounter CUDA runtime errors, RSH-SpMM achieves an average \textbf{2.10$\times$} speedup over HC-SpMM, which generally lags behind both pure CUDA-core methods such as RoDe and pure Tensor-Core designs such as Acc-SpMM. Similarly, MP-SpMM is built on SpTCs and relies on strict structured sparsity formats, which limits its applicability on general real-world matrices, resulting in an average \textbf{1.32$\times$} speedup for RSH-SpMM. In contrast, RSH-SpMM maintains stable performance across diverse sparsity patterns by employing finer-grained partitioning and explicitly addressing load imbalance.

\begin{figure}[t]
  \centering

  \begin{subfigure}{\linewidth}
    \centering
    \includegraphics[width=\linewidth]{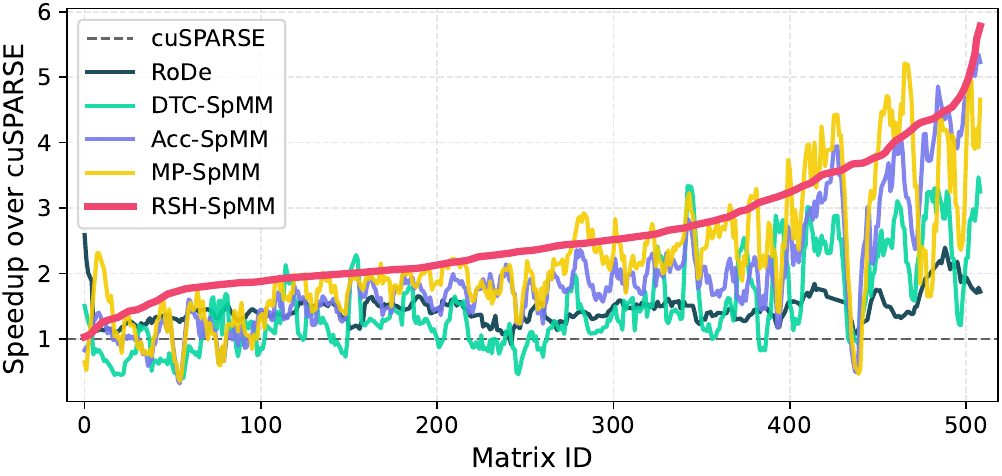}
    \caption{Normalized speedup over cuSPARSE on RTX~4090.}
    \label{fig:suitesparseresult_4090}
  \end{subfigure}
  \begin{subfigure}{\linewidth}
    \centering
    \includegraphics[width=\linewidth]{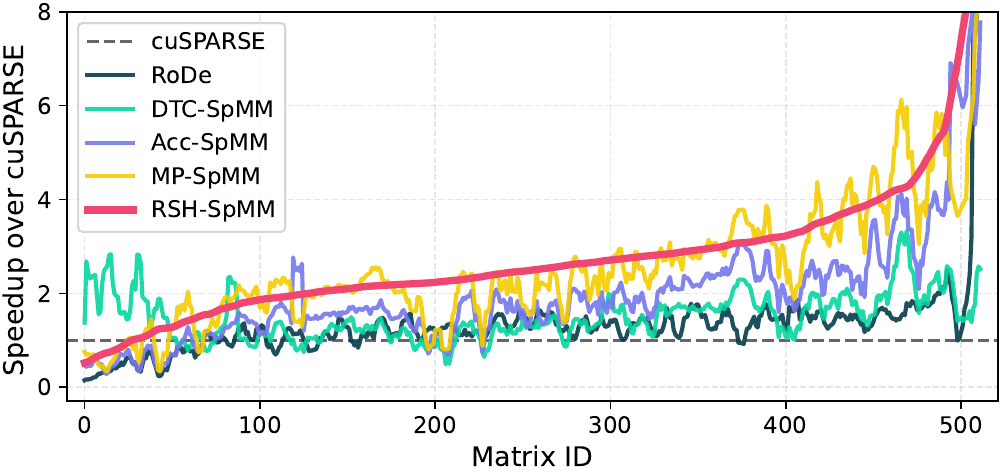}
    \caption{Normalized speedup over cuSPARSE on RTX~3090.}
    \label{fig:suitesparseresult_3090}
  \end{subfigure}

  \caption{Kernel-level SpMM performance on SuiteSparse matrices, reported as normalized speedup over cuSPARSE and sorted by RSH-SpMM speedup.}
  \label{fig:suitesparseresult}
\end{figure}

To assess robustness under broader sparsity diversity, we further evaluate RSH-SpMM on the SuiteSparse collection, with Figure~\ref{fig:suitesparseresult} summarizing kernel-level SpMM performance on RTX~4090 and RTX~3090. Across the collection, RSH-SpMM delivers consistent gains, achieving an average \textbf{3.21$\times$} speedup over cuSPARSE, with \textbf{over 80\%} of matrices falling in the \textbf{1.24$\times$--8.2$\times$} range. Compared with RoDe, RSH-SpMM achieves an average \textbf{2.04$\times$} improvement by leveraging Tensor Core throughput while preserving robustness on irregular structures. Relative to Tensor-Core approaches, it exceeds DTC-SpMM and Acc-SpMM by \textbf{1.91$\times$} and \textbf{1.61$\times$} on average, highlighting the difficulty of maintaining stable tile quality when window densities fluctuate. Against MP-SpMM, which relies on transforming matrices into strict 2:4 structured sparsity, RSH-SpMM provides an average \textbf{1.28$\times$} advantage by avoiding format constraints and preprocessing overhead. Overall, RSH-SpMM demonstrates the strongest and most consistent kernel-level performance across both real-world graph workloads and the SuiteSparse collection, validating the effectiveness of its adaptive tiling and hybrid execution strategy under real-world sparsity variations.

\subsection{Component Analysis}

\subsubsection{Compressed Format}

\begin{figure}[h]
  \centering
  \includegraphics[width=\linewidth]{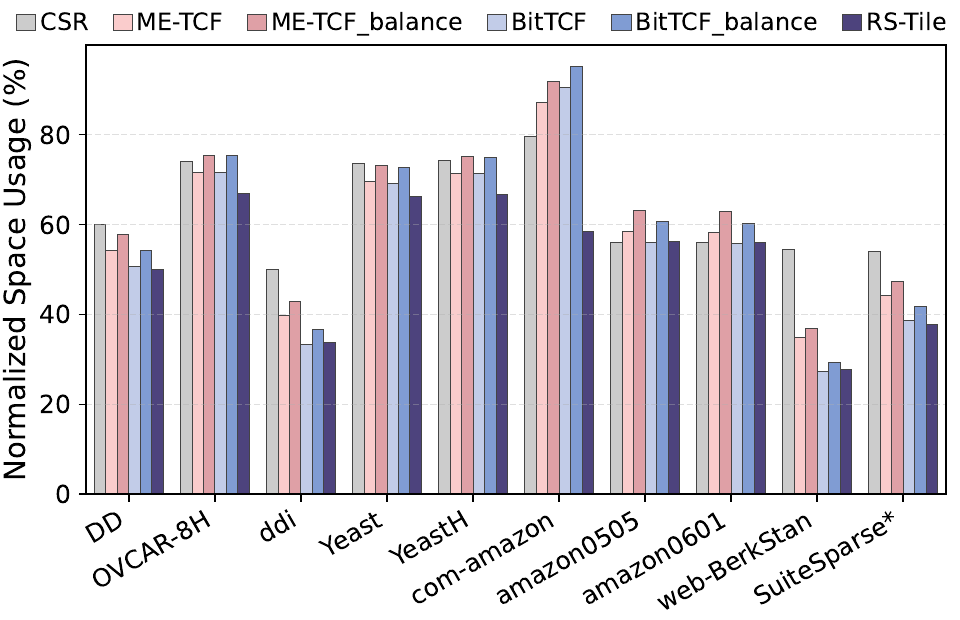}
  \caption{Normalized space usage (relative to COO) for all formats. For SuiteSparse, the value denotes the average over all matrices in the SuiteSparse collection; lower is better.}
   \label{fig:storage}
\end{figure}

\begin{figure*}[t]
  \centering
  \includegraphics[width=0.9\textwidth]{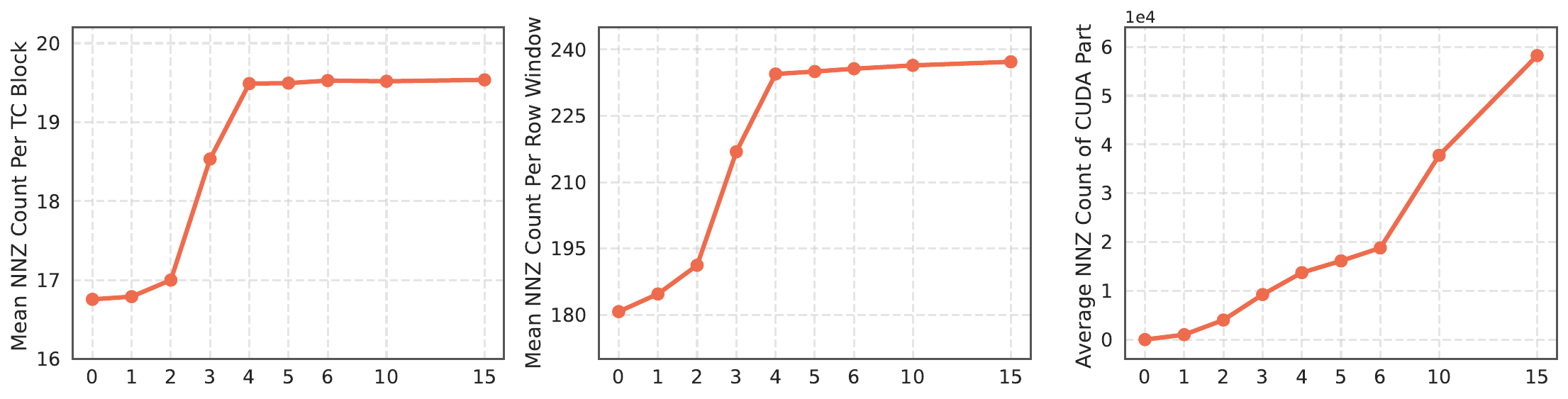}
  \caption{Effect of the row-nnz threshold on TC/CUDA partitioning. The horizontal axis shows the threshold applied to divide rows between TC and CUDA.}
   \label{fig:partition}
\end{figure*}

We compare the storage efficiency of RS-Tile with traditional sparse formats COO and CSR, as well as with the Tensor-Core-based formats used in DTC-SpMM and Acc-SpMM. For DTC-SpMM, we include both ME-TCF and its load-balanced variant ME-TCF\_balance; for Acc-SpMM, we include BitTCF together with its load-balanced variant BitTCF\_balance. All space costs are normalized to the COO baseline, as shown in Figure~\ref{fig:storage}.

Both ME-TCF and BitTCF reduce index overhead by condensing nonzero tiles, and BitTCF further compresses intra-tile column indices via bitmap encoding. However, once load balancing is enabled, these formats must maintain additional metadata, such as row-remapping tables or block-offset arrays, to track merged or split windows. This leads to notable storage inflation: the balanced variants of ME-TCF and BitTCF increase metadata size by \textbf{7.24\%} and \textbf{8.29\%}, respectively, with the overhead becoming especially pronounced for matrices exhibiting strong local density variation.

In contrast, RS-Tile integrates adaptive load balancing directly into its tile construction process, making the balancing mechanism inherently compatible with the storage layout. As a result, RS-Tile does not require duplicate index structures and avoids rewriting or augmenting row/block mappings after balancing. Consequently, the metadata footprint is determined entirely by the finalized tile layout and does not increase when load balancing is enabled.

Furthermore, the CUDA path in RS-Tile isolates extremely short or structurally incompatible rows from the Tensor-Core path, storing them in a compact fixed-row representation. This eliminates the need to decode short rows inside Tensor-Core tiles and reduces the indexing overhead on the Tensor-Core path accordingly.

Across all evaluated datasets,  RS-Tile consistently achieves more stable and compact storage usage. Compared with the balanced versions of ME-TCF and BitTCF, RS-Tile reduces metadata size by approximately \textbf{15.05\%} on average, directly lowering memory load/store overhead during the Tensor Core execution phase and improving sustained kernel throughput.

\subsubsection{Hybrid Kernel Analysis}

RSH-SpMM assigns the rows that are both extremely short and exhibit weak column-domain overlap with their surrounding row windows to the CUDA path, preventing such structurally isolated rows from disrupting the coherence of Tensor-Core windows. To evaluate the sensitivity of this strategy, we examine how adjusting the nnz threshold, which serves as a lightweight control knob for the degree of row removal, affects TC tile density. As shown in Figure~\ref{fig:partition}, increasing the threshold from 0 to 4 yields a clear improvement in TC tiling quality: the mean nnz per TC block increases from 16.7 to 19.5, and the mean nnz per row window rises from 174 to 233. Beyond this range, both metrics rapidly saturate, indicating that only a very small number of short or weakly correlated rows are responsible for most of the degradation in TC tile density. Within this low-threshold regime, the CUDA-assigned portion remains minimal, confirming that removing just these outlier rows is sufficient to stabilize TC tile structure without introducing noticeable additional cost.

When the threshold is increased further (beyond 6), TC tile density remains nearly unchanged while the CUDA workload grows sharply. In such cases, a significant amount of work is shifted to CUDA cores, whose lower arithmetic throughput and reduced parallel efficiency can substantially limit end-to-end performance. Therefore, RSH-SpMM deliberately restricts the threshold to a small range, typically within the plateau region indicated above, and determines its exact value adaptively based on matrix size and overall sparsity. This strategy improves TC tile regularity while preventing the CUDA path from becoming a performance bottleneck, enabling stable and efficient hybrid execution.

\begin{figure}[h]
  \centering
  \includegraphics[width=\linewidth]{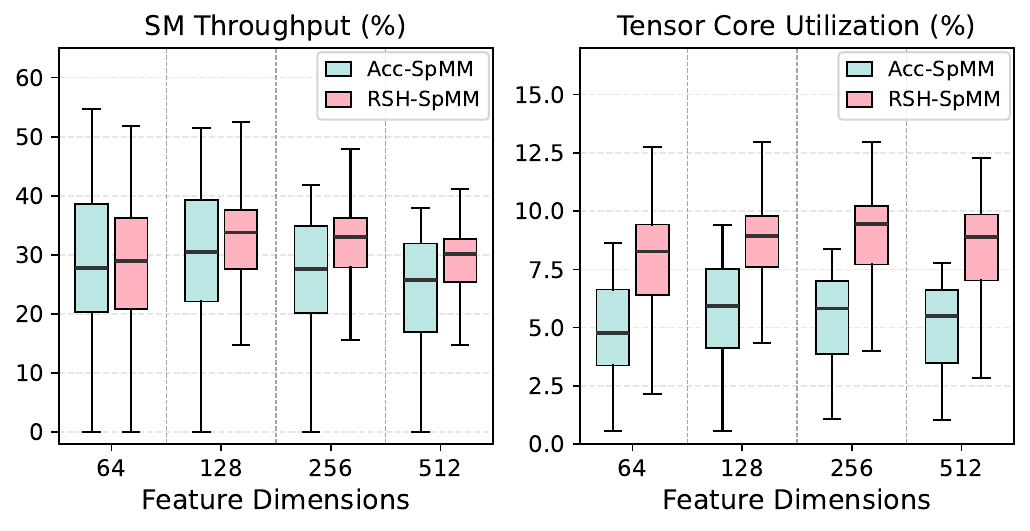}
  \caption{SM throughput and Tensor Core utilization comparison for Acc-SpMM and RSH-SpMM.}
   \label{fig:pipeline-prof}
\end{figure}

We further analyze the behavior of the hybrid kernel through GPU profiling and a fine-grained breakdown of kernel configurations. Figure~\ref{fig:pipeline-prof} compares RSH-SpMM with Acc-SpMM in terms of SM compute throughput and Tensor Core utilization, averaged across all datasets and feature dimensions. Across both metrics, RSH-SpMM exhibits a consistently higher and more concentrated distribution. Specifically, its SM throughput improves from a median of 28\% to 33\% (an increase of 18\%). Tensor Core utilization exhibits an even larger gain: the average utilization rises from 5.6\% in Acc-SpMM to 8.8\% under RSH-SpMM. This suggests that RSH-SpMM keeps the Tensor Core MMA pipeline more continuously occupied, with fewer idle cycles and a higher level of effective pipeline saturation.

Taken together, these results indicate that RSH-SpMM’s hybrid kernel sustains a more continuously utilized Tensor Core pipeline through selective row removal and balanced, memory-efficient TC execution. This synergy enables consistently higher performance than prior Tensor-Core-based or hybrid SpMM methods.

\subsubsection{Reordering Effects}

\begin{figure}[h]
  \centering
  \includegraphics[width=0.95\linewidth]{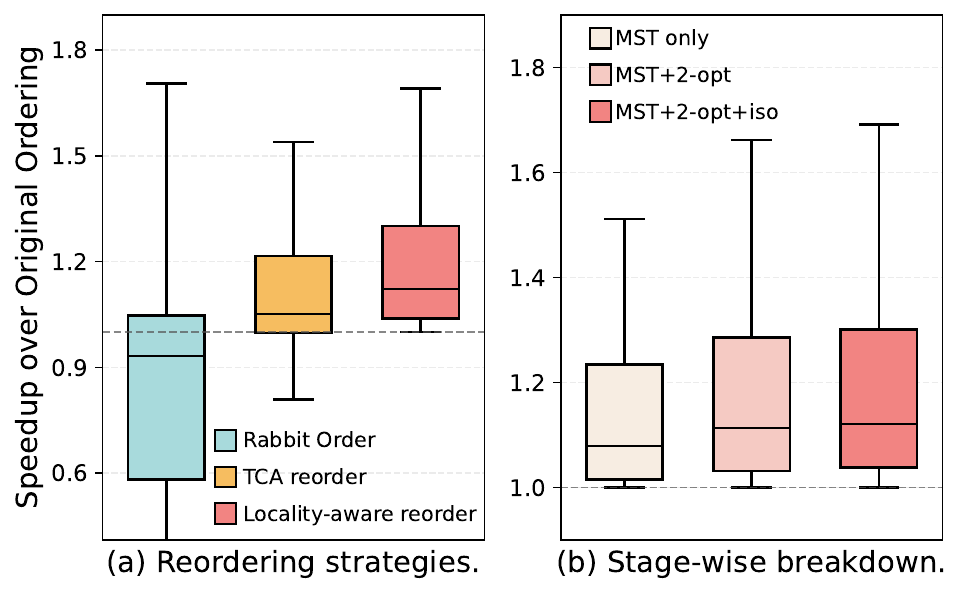}
  \caption{Effect of different reordering strategies on kernel-level SpMM performance, reported as relative speedup over the original ordering. }
   \label{fig:reorderresult}
\end{figure}

We compare our locality-aware reordering with two widely used baselines for high-performance sparse kernels: the Rabbit Order~\cite{rabbit} and the TCU-cache-aware (TCA) reordering used in DTC-SpMM. Our evaluation measures kernel-level SpMM speedups relative to the original input ordering across all evaluated datasets. Rabbit provides limited improvement, yielding an average speedup of \textbf{0.87$\times$}, while TCA delivers a more stable enhancement with an average of \textbf{1.15$\times$}. In contrast, our method achieves substantially higher gains on the \emph{full SpMM kernel}, reaching an average of \textbf{1.25$\times$} and up to \textbf{1.7$\times$} on certain workloads, as shown in Figure~\ref{fig:reorderresult}(a).

To understand where these benefits originate, we examine the contribution of each stage of our pipeline (Figure~\ref{fig:reorderresult}(b)). The \emph{MST-based ordering} provides the dominant improvement, raising performance to \textbf{1.19$\times$}. The subsequent \emph{2-opt refinement} further enhances locality, increasing the speedup to \textbf{1.22$\times$}. The final \emph{isolation-aware adjustment} resolves structurally incoherent rows that would otherwise reduce spatial locality, raising performance to reach \textbf{1.25$\times$}.

These three stages together yield a locality-friendly row permutation that strengthens locality, reduces memory stalls, and consistently improves SpMM kernel performance over existing methods.

\subsection{Case Study: End-to-end Training Performance for GCN}

To assess the practical impact of RSH-SpMM in real GNN workloads, we integrate it into the forward computation of a 6-layer Graph Convolutional Network (GCN)~\cite{gcn} implemented in PyTorch. Each GCN layer performs the standard sparse-dense multiplication
\begin{equation}
H^{(\ell+1)} = \sigma \big( A H^{(\ell)} W^{(\ell)} \big),
\end{equation}
where $A$ is the sparse adjacency matrix, and the multiplication $A H^{(\ell)}$ forms the SpMM kernel that dominates the sparse computation cost. Replacing this operation directly affects the end-to-end training time of GNN models.

We evaluate the training performance on four representative graph datasets (web-BerkStan, com-amazon, DD, and amazon0601) on RTX~4090 GPU. Each model is trained for 500 epochs using two hidden dimensions, $d=64$ and $d=256$, and we compare the end-to-end runtime when adopting RSH-SpMM, PyTorch Geometric (PyG)~\cite{pyg}, cuSPARSE, DTC-SpMM, and TC-GNN as the SpMM backend, respectively.

\begin{figure}[h]
  \centering
  \includegraphics[width=\linewidth]{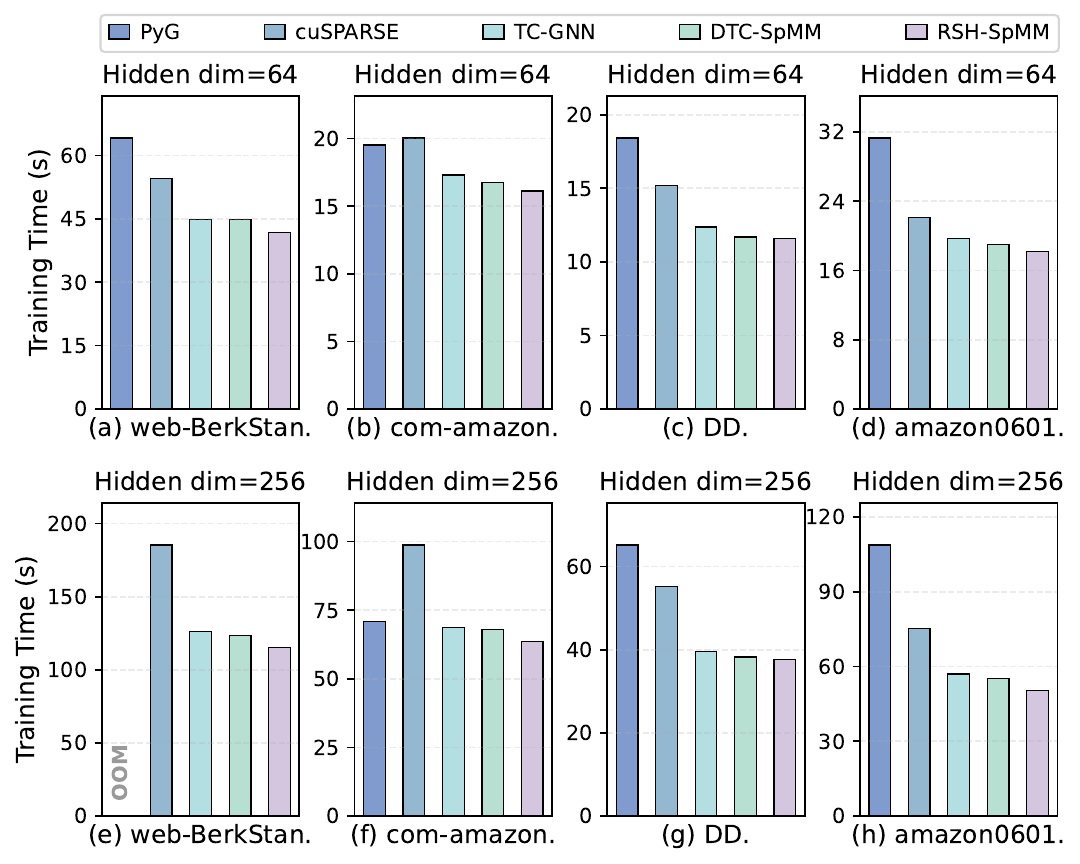}
  \caption{End-to-end GCN training on an RTX~4090, running for 500 epochs. OOM denotes out-of-memory.}
   \label{fig:end2end}
\end{figure}

RSH-SpMM achieves the lowest overall training time of \textbf{44.32\,s} when averaged across the four datasets and the two feature dimensions, as summarized in Figure~\ref{fig:end2end}. In comparison, DTC-SpMM, TC-GNN, PyG, and cuSPARSE require 47.10\,s, 48.23\,s, 54.05\,s, and 65.82\,s, respectively, corresponding to average speedups of \textbf{1.06$\times$}, \textbf{1.09$\times$}, \textbf{1.22$\times$}, and \textbf{1.49$\times$} over these baselines.

These results indicate that the row-structured tiling and hybrid execution strategy of RSH-SpMM effectively translates kernel-level improvements into end-to-end GCN training gains. By accelerating the SpMM operation in every layer while maintaining robustness under irregular sparsity, RSH-SpMM consistently outperforms prior Tensor-Core-accelerated designs (TC-GNN and DTC-SpMM) as well as framework-level implementations (PyG and cuSPARSE), demonstrating its effectiveness across diverse graph datasets.

\section{Related Work}

Research on GPU-based SpMM spans a wide range of kernel designs and sparse representations, including both general-purpose implementations such as cuSPARSE~\cite{nvidia_cusparse_2024} and specialized systems tailored to different sparsity patterns. Prior work can be broadly grouped into CUDA-core approaches, Tensor-Core approaches, and hybrid designs combining the two. CUDA-core methods such as Sputnik~\cite{sputnik}, GE-SpMM~\cite{ge-spmm}, ASpT~\cite{aspt}, and RoDe~\cite{rode} focus on improving warp cooperation, coalescing, and row-level scheduling for fine-grained irregularity. Tensor-Core approaches, such as TC-GNN~\cite{tc-gnn}, DTC-SpMM~\cite{dtc-spmm}, FlashSparse~\cite{flashsparse}, Acc-SpMM~\cite{acc-spmm}, SMaT~\cite{smat}, and Voltrix~\cite{voltrix}, restructure sparse matrices into dense-aligned tiles suitable for MMA execution. Others such as MP-SpMM~\cite{mp-spmm} and JigSaw~\cite{jigsaw} leverage structured sparsity to activate Sparse Tensor Cores. Hybrid approaches such as HC-SpMM~\cite{hc-spmm}, HR-SpMM~\cite{hr-spmm}, and BRP-SpMM~\cite{brp-spmm} combine CUDA cores and Tensor Cores by routing different rows or regions based on sparsity statistics or per-row heuristics. Together, these studies reflect the diverse strategies explored for accelerating SpMM on modern GPUs.

Reordering techniques are widely used to enhance spatial locality and expose denser substructures in sparse operators. Reordering methods, including community-based approaches~\cite{communityreorder}, graph-centric techniques such as Rabbit Order~\cite{rabbit} and SlashBurn~\cite{slashburn}, and affinity-preserving row permutations like Groot~\cite{Groot} and Graphite~\cite{graphite}, construct matrix or graph permutations that improve locality by reducing working-set size and increasing structural coherence, thereby benefiting downstream sparse kernels including SpMM. These techniques primarily focus on preprocessing the sparsity pattern to create more locality-friendly matrix layouts rather than redesigning the execution pipeline itself.

Sparse computation is also extensively studied across modern graph learning and Large Language Model (LLM) systems.
Frameworks such as GNNAdvisor~\cite{gnnadvisor}, PckGNN~\cite{pckgnn}, and GNNPilot~\cite{gnnpilot} optimize dataflow, operator scheduling, and kernel selection to better utilize GPU resources under diverse graph workloads, while fused operators such as FusedMM~\cite{fusedmm} improve data reuse across SDDMM-SpMM pipelines, where SDDMM denotes sampled dense-dense matrix multiplication. In large-language-model inference, systems including FlashLLM~\cite{flashllm}, SpInfer~\cite{spinfer}, and GeneralSparse~\cite{generalsparse} exploit static or dynamic sparsity to accelerate attention and MLP operations. Beyond deep learning, SpMM likewise powers linear-algebraic graph analytics such as GraphBLAS~\cite{graphblas}, where graph tasks are realized via sparse matrix kernels. These system-level efforts demonstrate the broad importance of efficient sparse operators and motivate kernel designs that remain effective under highly irregular sparsity.

\section{Conclusion}

In this paper, we presented \textbf{RSH-SpMM}, a row-structured hybrid SpMM kernel designed for modern GPUs. To address the substantial irregularity in real-world sparse matrices, such as variations in row sizes and local structures, we developed a unified framework that integrates optimizations across representation, scheduling, and execution. Extensive experiments demonstrate that RSH-SpMM achieves significant and stable performance improvements over state-of-the-art GPU SpMM baselines.

\bibliographystyle{ACM-Reference-Format}
\bibliography{ref}

\end{document}